\documentclass[a4paper]{jpconf}
\usepackage{graphicx}
\begin{document}
\title{Heavy quark and $J$/$\psi$ production at RHIC/PHENIX}

\author{Tsuguchika Tabaru for the PHENIX Collaboration}

\address{
  RIKEN BNL Research Center,
  Building 510A, Brookhaven National Laboratory, Upton, NY 11973, USA
}

\ead{tsugu@bnl.gov}

\begin{abstract}
Single electrons from open heavy quarks and di-leptons from $J$/$\psi$ mesons
have been studied systematically at RHIC/PHENIX
using data from $p$+$p$, $d$+Au and Au+Au collisions at
$\sqrt{s_{NN}}$=62.4 GeV, 130 GeV and 200 GeV. From the single electron study,
the charm quark yield is found to scale with the number of binary collisions.
This scaling has recently been confirmed using Au+Au collisions at 62.4 GeV.
A new analysis shows that a high $p_{\rm T}$ suppression of single electrons
is observed in Au+Au collisions at 200 GeV.
This suppression suggests that heavy quarks lose significant energy
in the medium.
A weak rapidity dependence is seen in  $J$/$\psi$ yields from $d$+Au data,
which can be interpreted as a cold matter effect.
We report also the results of a measurement of the spin alignment of the 
$J$/$\psi$ in the $p_{\rm T}$ range around 1.5 GeV/$c$.
\end{abstract}

\section{Introduction}
There is growing evidence that a new state of hot and/or dense matter
is formed in high energy heavy ion collisions at RHIC \cite{PHENIX White Paper}.
This new state of matter is thought to be made of de-confined quarks and gluons,
and it is expected to behave differently from normal nuclear matter.
Heavy quarks (charm and bottom) and quarkonia are produced via hard processes
in the early stage of the collisions,
and they travel through the medium of partons.
Due to their heavy mass, heavy quarks can experience interaction
with the medium in a different way from light quarks \cite{dead cone}.
Therefore, measurements of open heavy quark yields in heavy ion collisions
provide new information on the nature of the medium created at RHIC.
Also, it is predicted that the $J$/$\psi$ yield will be either
suppressed \cite{J/psi suppression}
or enhanced \cite{J/psi enhancement}
in the presence of de-confined quarks and gluons.

In high energy nuclear collisions,
a large fraction of electrons with transverse momentum ($p_{\rm T}$)
larger than 1 GeV/$c$ comes from decays of heavy quarks.
Thus, one can measure heavy quark production via single electrons.
PHENIX has measured open heavy quark yields using single electrons,
and $J$/$\psi$ yields using both di-electron and di-muon pairs, allowing a 
systematic study of the effects of nuclear matter on these heavy quark signals.
In this article, the recent results of the heavy quark data at PHENIX
are presented.

\section{Open heavy quark measurement}
The open heavy quark yields were studied through their electron decays.
The electrons were measured in two central arms \cite{PHENIX detector},
each covering $\Delta \phi$=$\pi$/2 at mid-pseudo-rapidity
($\left| \eta \right|$$<$0.35).
The charged tracks were reconstructed with drift chambers and pad chambers.
In each arm the electrons were identified by  a Ring Imaging \v Cerenkov 
counter (RICH)
and an ElectroMagnetic Calorimeter (EMCal)
in the $p_{\rm T}$ range of 0.5 to 5 GeV/$c$.

Yields of electrons coming from
heavy quark decays are extracted by subtracting background electrons.
The main components of the background electron yield, called ``photonic electrons"
come from the Dalitz decays of neutral pions and eta mesons, and from gamma
conversions.
All other electrons, including those from heavy quark decays,
are called ``non-photonic electrons".
The backgrounds in non-photonic electrons are small,
and come from di-electron decays and semi-leptonic decays of other mesons.
PHENIX evaluates the backgrounds mainly by the following two methods.
In the first method, the ``converter subtraction" method,
special data sets were taken with
a thin brass cylinder (radiation length of 1.7\%)
installed around the collision vertex position.
Since the brass converter increases only the yield of conversion electrons,
the yield and $p_{\rm T}$ shapes of photonic electrons can be evaluated
experimentally and precisely, especially in the low $p_{\rm T}$ region.
The advantage of this method is that the systematic error
in the result is small.
The disadvantage is that the statistical error is dominated by
the data taken with the converter, about 10\% of all the data.
In the second method, the ``cocktail subtraction" method,
all of the background electron yields are evaluated by simulation
using as input the meson and photon distributions measured by the PHENIX detector.
This method is good to study signal electrons at high $p_{\rm T}$,
where signal to background ratio is larger than unity.
The results of the cocktail subtraction measurement are confirmed by those of
the converter subtraction method.

\begin{figure}[h]
\begin{minipage}{18pc}
\includegraphics[width=18pc]{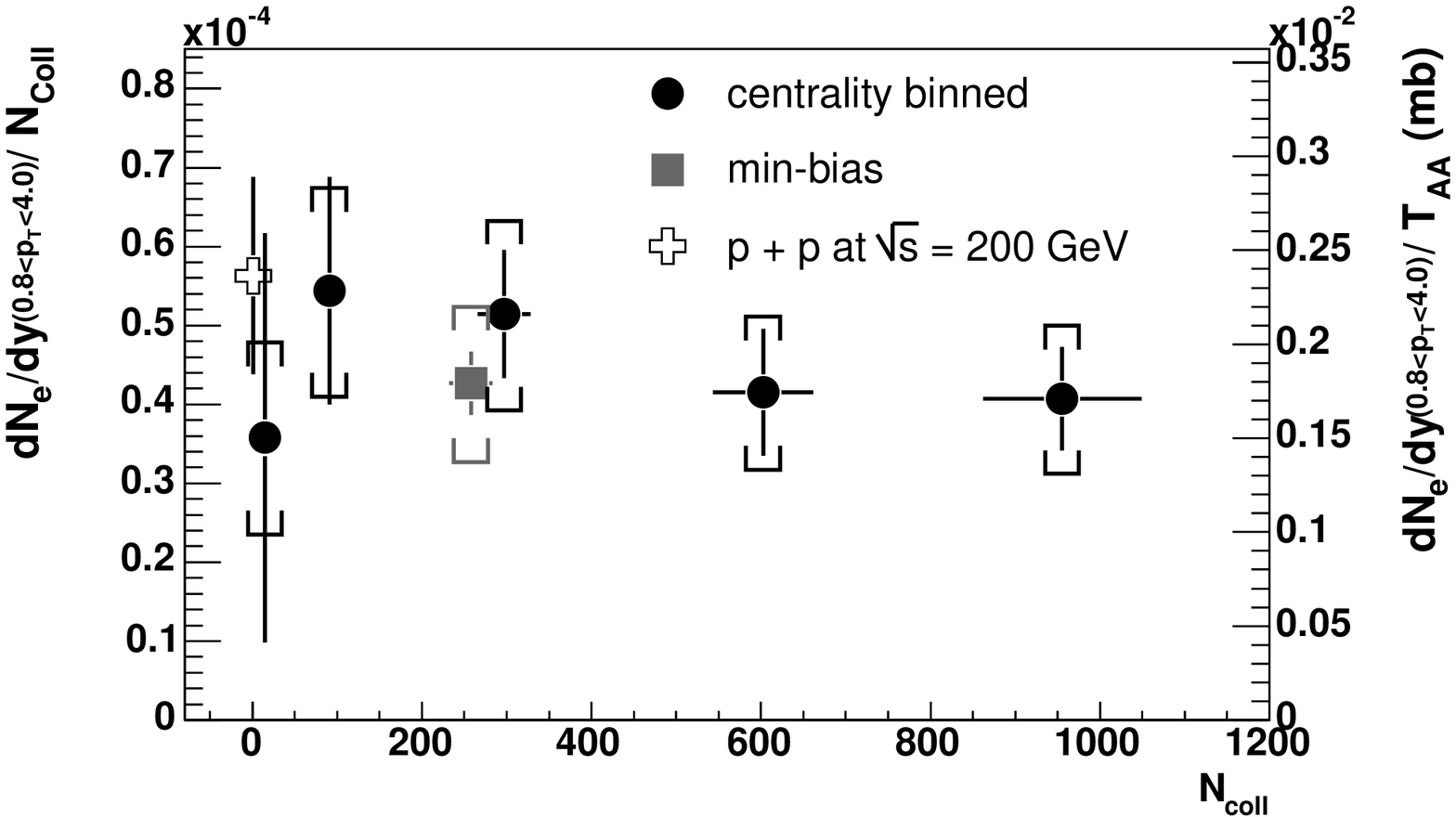}
\caption{\label{Fig.1}
  Non-photonic electron yield (0.8$<$$p_{\rm T}$$<$4.0 GeV/$c$)
  measured in Au+Au collisions at $\sqrt{s_{NN}}$=200 GeV scaled with
  $N_{\rm coll}$ (left-hand scale) as a function of centrality given by
  $N_{\rm coll}$.
  This electron yield translates to the electron cross section per
  $NN$ collision in the above $p_{\rm T}$ range (right-hand scale).
}
\end{minipage}\hspace{2pc}%
\begin{minipage}{18pc}
\includegraphics[width=18pc]{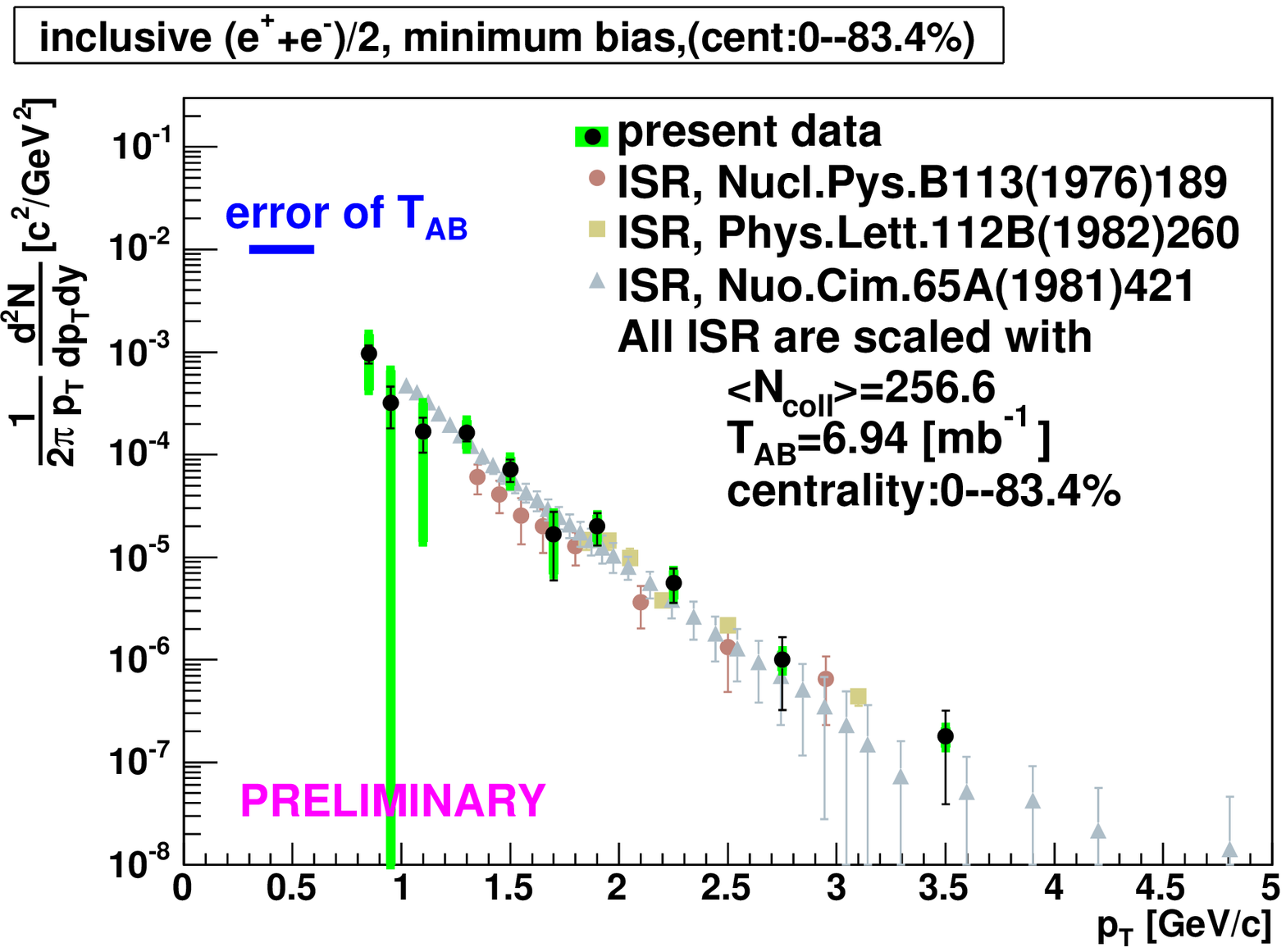}
\caption{\label{Fig.2}
  Single electron $p_{\rm T}$ spectra in Au+Au collisions (new data)
  and the ISR $p$+$p$ data \cite{ISR data 1, ISR data 2, ISR data 3}
  at $\sqrt{s_{NN}}$=62 GeV.
  The ISR data are scaled by $N_{\rm coll}$.
}
\end{minipage} 
\end{figure}

To study the behavior of heavy quarks in the new state of matter systematically,
various data were collected in $p$+$p$, $d$+Au,
and Au+Au \cite{AuAu 200 GeV single-e} collisions at $\sqrt{s_{NN}}$=200 GeV,
and Au+Au at 62 GeV and 130 GeV \cite{AuAu 130 GeV single-e}.
Comparison between $p$+$p$ and $d$+Au data \cite{dAu 200 GeV single-e},
shows that
the $p_{\rm T}$ distribution of single electrons scales with
the number of binary collisions ($N_{\rm coll}$),
within the uncertainty of the measurement.
The total yield of single electrons is also found to be proportional to
$N_{\rm coll}$ in Au+Au collisions.
Figure~\ref{Fig.1} shows the Au+Au results \cite{AuAu 200 GeV single-e},
with the signal electron yield
normalized by $N_{\rm coll}$ as a function of $N_{\rm coll}$.
This was also studied at 62.4 GeV.
Figure~\ref{Fig.2} shows the new result,
the $p_{\rm T}$ spectra of signal electrons in Au+Au collisions at 62.4 GeV.
There are also plotted
the ISR single electron data taken in $p$+$p$ collisions at the same energy,
scaled by $N_{\rm coll}$ \cite{ISR data 1, ISR data 2, ISR data 3}.
The plot shows that the single electron yield scales with $N_{\rm coll}$.
These analyses were done using the converter subtraction method.

\begin{figure}[h]
\includegraphics[width=21pc]{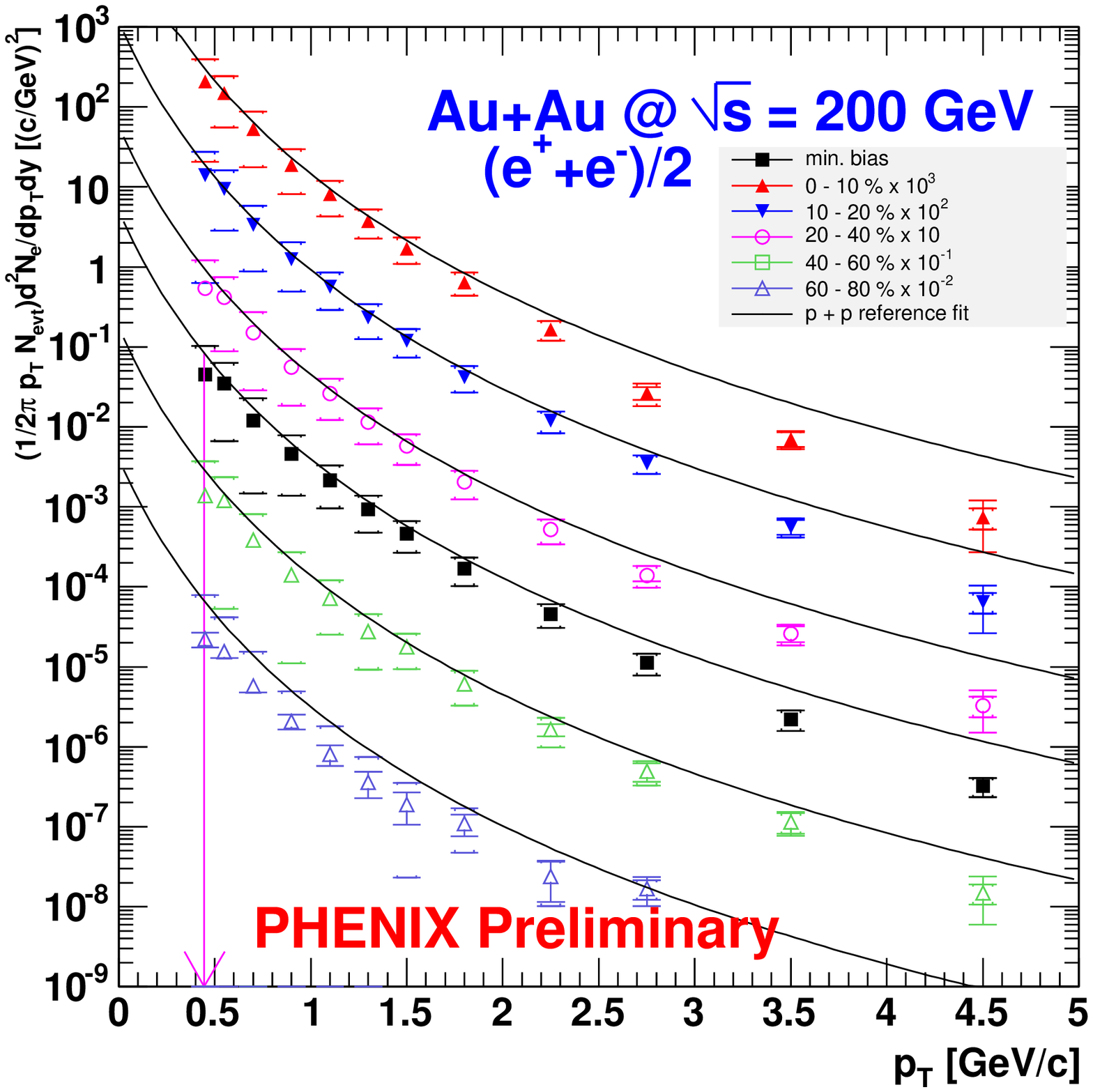}
\includegraphics[width=18pc]{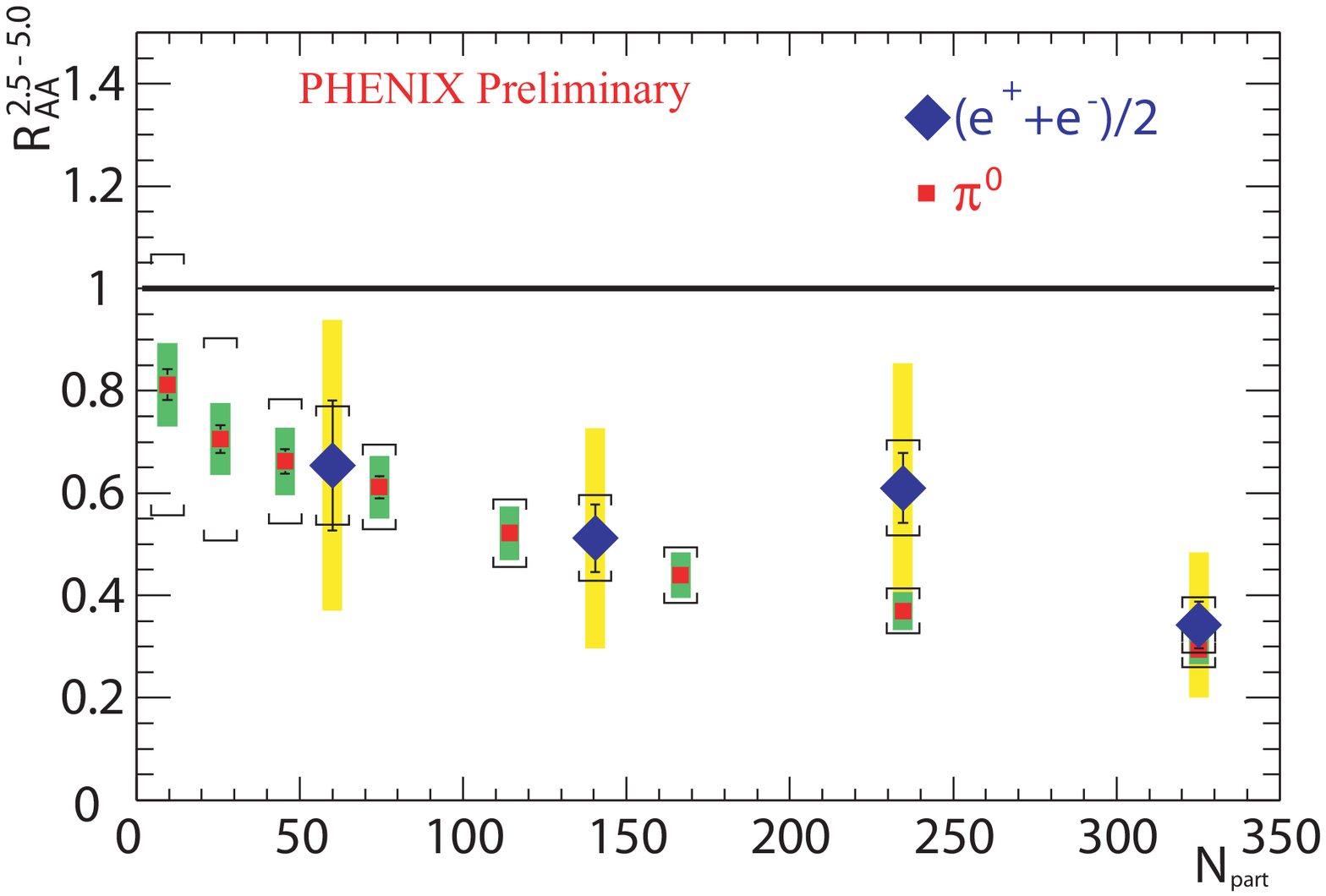}
\caption{\label{Fig.3}
  Left: single electron $p_{\rm T}$ spectra for each centrality bin
  in Au+Au collisions at $\sqrt{s_{NN}}$=200 GeV (points),
  compared with the $p$+$p$ result scaled by number of binary collisions
  (solid lines).
  Right: nuclear modification factor ($R_{AA}$) of single electron (diamond)
  in 2.5$<$$p_{\rm T}$$<$5.0 GeV as a function of the number of participants.
  Those of neutral pions are also plotted by square points as reference data.
}
\end{figure}

Figure~\ref{Fig.3} shows new results, namely the $p_{\rm T}$ distribution of
single electrons
from heavy quark decays in Au+Au collisions at $\sqrt{s_{NN}}$=200 GeV,
and the nuclear modification factor in 2.5$<$$p_{\rm T}$$<$5.0 GeV
as a function of the number of participants.
This analysis uses the same data set as that of Fig.~\ref{Fig.1}.
The statistics are greatly improved because the cocktail subtraction method 
was used.
The new high statistics result clearly shows that the single electron
spectra are suprresed at high $p_{\rm T}$.
This suppression suggests that heavy quarks lose
significant energy in the media.
The elliptic flow of charm mesons is also
measured \cite{AuAu 200 GeV single-e v2}.
The measured eliptic flow of heavy flavor electron is nonzero with a
90 \% confidence level.
The new Au+Au data taken in 2004 should clarify these things
because of their much higher statistics.

\section{$J$/$\psi$ measurement}
\begin{figure}[h]
\begin{minipage}{18pc}
\includegraphics[width=18pc]{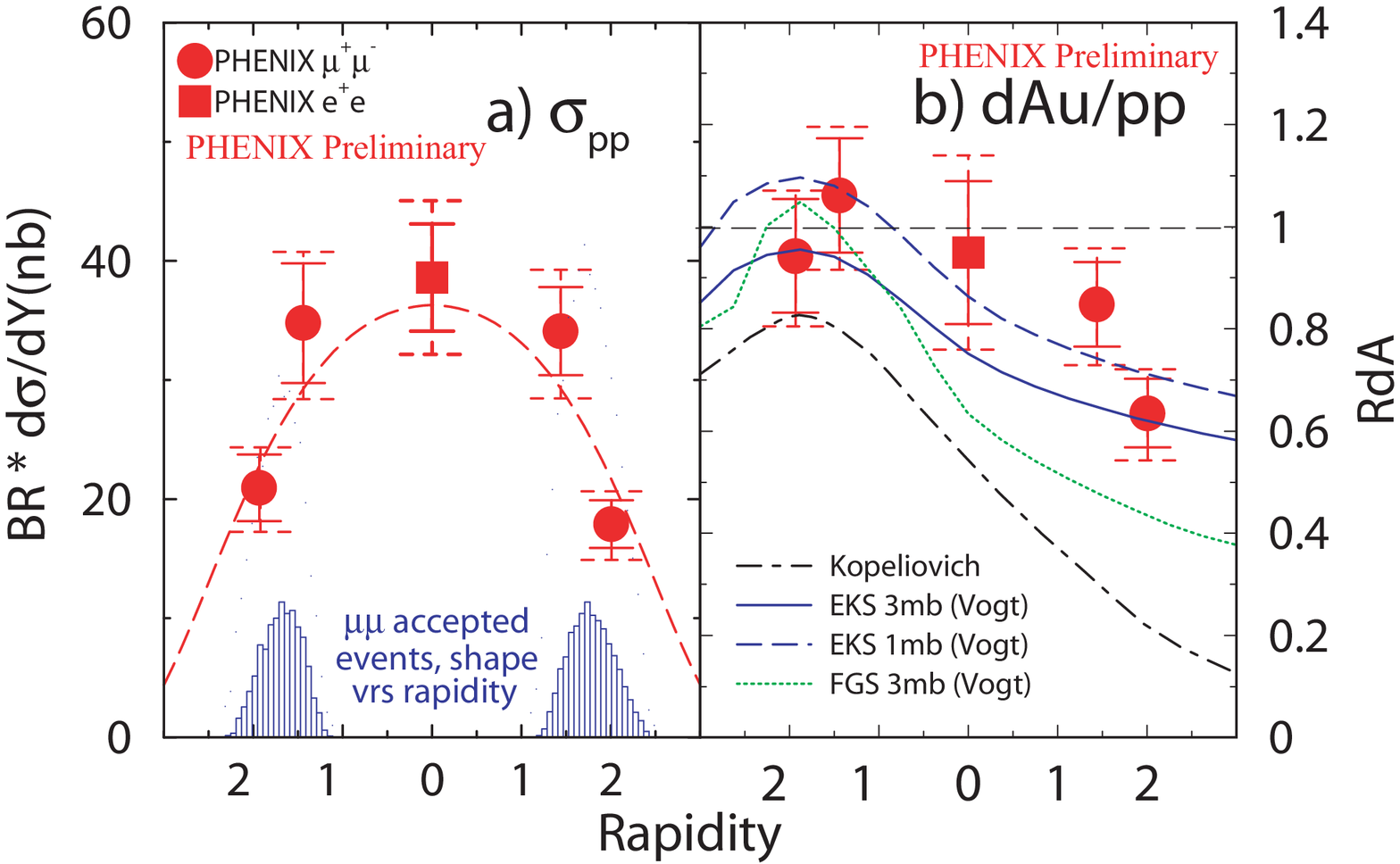}
\caption{\label{Fig.4}
  The left panel shows the $J$/$\psi$ production cross section
  as a function of rapidity
  in $p$+$p$ collisions at $\sqrt{s_{NN}}$=200 GeV.
  The right panel shows the nuclear modification factor for $d$+Au collisions
  ($R_{dA}$) as a function of rapidity.
}
\end{minipage}\hspace{2pc}%
\begin{minipage}{18pc}
\includegraphics[width=18pc]{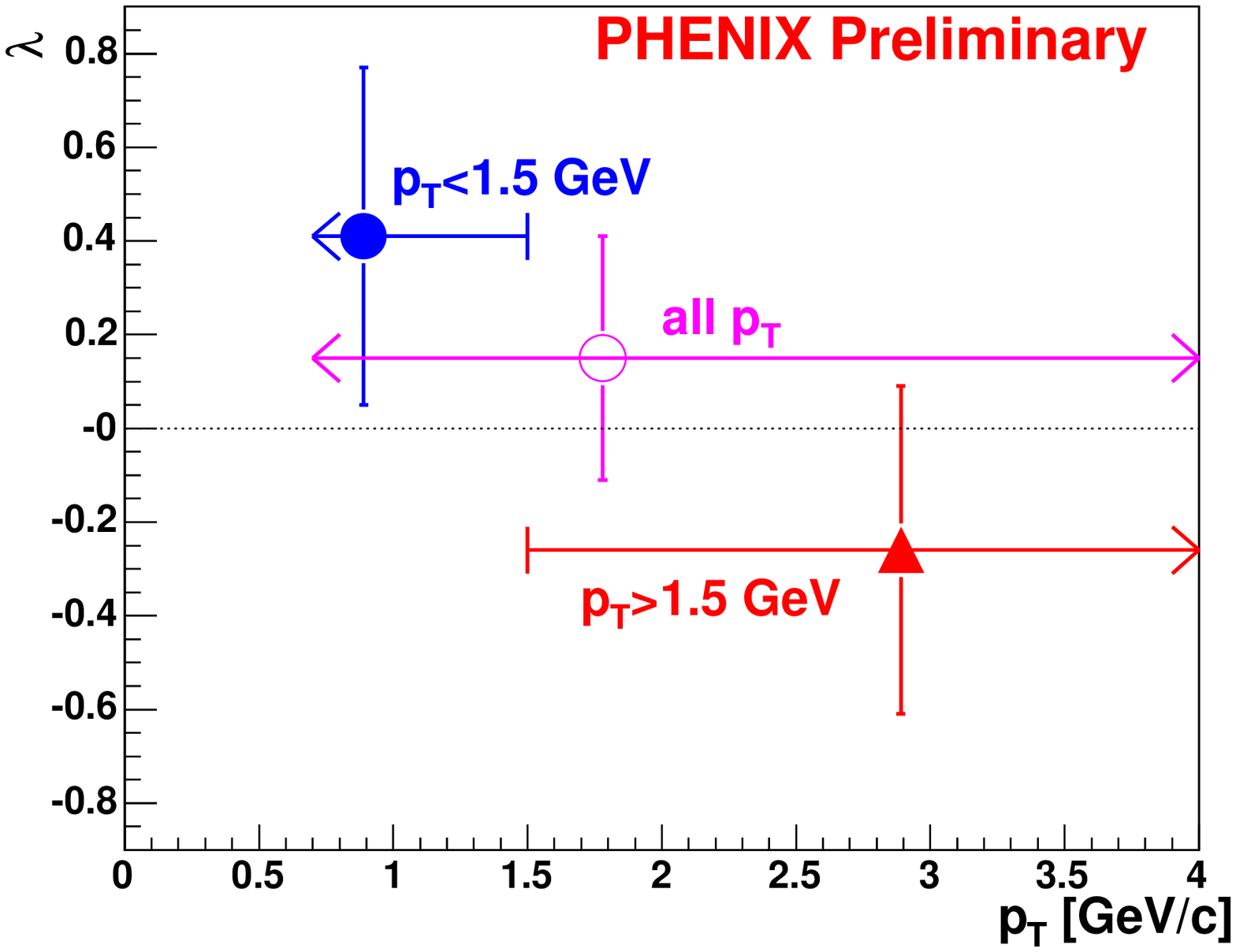}
\caption{\label{Fig.5}
  The $J$/$\psi$ polarization parameter as a function of $p_{\rm T}$
  in $d$+Au collisions at $\sqrt{s_{NN}}$=200 GeV.
  The filled circle and the triangle points show the $\lambda$
  in $p_{\rm T}$ below and above 1.5 GeV/$c$, respectively.
  The open circle shows the polarization parameter for all $p_{\rm T}$.
}
\end{minipage} 
\end{figure}

The $J$/$\psi$ meson yields were measured via the di-electron decay channel
using the central arms
and via the di-muon decay channel using the muon arms
at forward (1.2$<$$\eta$$<$2.4)
and backward pseudo-rapidity ($-2.2$$<$$\eta$$<$$-1.2$).
The production cross sections were measured in $p$+$p$ \cite{pp 200 GeV J/psi}
(Fig.~\ref{Fig.4}-(a)), $d$+Au and Au+Au \cite{AuAu 200 GeV J/psi} collisions.
The nuclear modification factor for $d$+Au, the ratio of the binary collision
scaled $d$+Au cross section to the corresponding $p$+$p$ cross section, is
shown versus rapidity in (Fig.~\ref{Fig.4}-(b)).
There is a weak rapidity dependence that suggests effects of nuclear 
absorption or nuclear (anti-) shadowing.
We expect about 600 $J$/$\psi \to e^+ e^-$ events and several thousand 
$J$/$\psi \to \mu^+ \mu^-$ events from the Au+Au data set taken 
in 2004, which is still being analyzed. This should allow us to make a plot 
similar to (Fig.~\ref{Fig.4}-(b)) for Au+Au.

The production mechanism of $J$/$\psi$ mesons was also studied
with $J$/$\psi$ spin alignment in $d$+Au data at mid-rapidity
in the $p_{\rm T}$ around 1.5 GeV/$c$.
The spin alignment was measured by the positron angular distribution,
$1 + \lambda \cos^2\theta$,
where $\theta$ is the angle between the positron momentum
and the momentum direction of the $J$/$\psi$ in the rest frame.
Figure~\ref{Fig.5} shows the measured $\lambda$ as a function of $p_{\rm T}$.

\section{Summary}
Open heavy quark yields were measured by PHENIX through their
decay electrons at $\sqrt{s_{NN}}$=200 GeV in $p$+$p$, $d$+Au
and Au+Au collisions, and at 62.4 GeV and 130 GeV in Au+Au collisions.
The electrons $p_{\rm T}$ spectra from open heavy flavor in $d$+Au 
scale with $N_{\rm coll}$ at mid rapidity.
In Au+Au collisions at 200 GeV, the total yield of single electrons
scales with the number of binary collisions,
while the single electron yields are suppressed at high $p_{\rm T}$.

The $J$/$\psi$ meson production cross sections were measured
in $p$+$p$, $d$+Au and Au+Au collisions at 200 GeV.
A weak rapidity dependence is seen in the $d$+Au nuclear modification factor.
No significant spin alignment of $J$/$\psi$ is seen at the mid-rapidity
in the $p_{\rm T}$ range around 1.5 GeV/$c$.

\end{document}